\documentclass[12pt]{article}

\usepackage{amsmath}
\usepackage{pdflscape}
\usepackage{color}
\usepackage{amssymb}
\usepackage{graphicx}
\usepackage{subfigure}
\usepackage{hyperref}
\usepackage{mathrsfs}
\usepackage{latexsym}
\usepackage{amsfonts}
\usepackage{amsmath}
\usepackage{esint}
\usepackage{amsxtra}
\usepackage{longtable}
\usepackage{rotating}
\usepackage{multirow}
\usepackage[titletoc,title]{appendix}
\usepackage{wrapfig}

\newdimen\mathindent
\mathindent\leftmargini

\def \ep1{\epsilon_1}
\def \ep2{\epsilon_2}
\def \m{\mbox}

\def \be{\begin{equation}}
\def \ee{\end{equation}}
\def\beq{\begin{eqnarray}}
\def\eeq{\end{eqnarray}}
\def \ba{\begin{array}}
\def \ea{\end{array}}

\def \f{\frac}

\def \p{\partial}

\def \sn{\mbox{sn}}
\def \cn{\mbox{cn}}
\def \dn{\mbox{dn}}
\def \ep{\epsilon}

\begin{document}
\vspace*{-.6in}
\thispagestyle{empty}
\baselineskip = 18pt

\vspace{.5in}
\vspace{.5in}
{\LARGE
\begin{center}
A new treatment for some periodic Schr\"{o}dinger operators I: the eigenvalue
\end{center}}

\vspace{1.0cm}

\begin{center}
\renewcommand{\thefootnote}{\fnsymbol{footnote}}
Wei He\footnote[1]{weihephys@foxmail.com}
\setcounter{footnote}{0}
\vspace{1.0cm}

\em{Instituto de F\'isica Te\'orica, Universidade Estadual Paulista,\\
Barra Funda, 01140-070, S\~{a}o Paulo, SP, Brazil}

\end{center}
\vspace{1.0cm}

\begin{center}
\textbf{Abstract}
\end{center}
\begin{quotation}
\noindent  We study the problem of how the Floquet property manifests for periodic Schr\"{o}dinger operators which are known to have multiple of asymptotic spectral solutions.
The main conclusions are made for elliptic potentials, we demonstrate that for each period of the elliptic function there is a relation about the Floquet exponent and the monodromy of wave function. Among them there are two relations not explained by the classical Floquet theory.
These relations produce both old and new asymptotic solutions consistent with results already known.
\end{quotation}




\pagenumbering{arabic}

\newpage

\tableofcontents

\vspace{1cm}

\section{Introduction}

Consider the following 1-dimensional stationary Schr\"{o}dinger
equation with periodic potential, i.e. a second order periodic ordinary
differential equation
\begin{equation} (\p_x^2-u(x))\psi=\lambda\psi,\qquad
u(x)=u(x+T).\label{schro-eq}
\end{equation}
It is applied in many areas, from
celestial mechanics to accelerator physics and quantum mechanics. There
is a large amount of literatures about the linear problem with
periodic coefficient, such as Refs.~\cite{Whittaker-Watson, mclachlan, Erdelyi3, Arscott1964,
Magnus-Winkler, Eastham1973, Wang-Guo, MullerKirstenQUANT, ErdelyiAsym, NIST} and references therein. In this paper we
focus on the particular aspect about asymptotic solution for the
spectrum $\lambda$. By ``{\em asymptotic solution}" we mean a solution expanded as an asymptotic series controlled by a small/large parameter \cite{ErdelyiAsym}.
The parameter space of the equation consists of $\lambda$ and the coupling strength of $u(x)$ collectively denoted by $g$.
A different asymptotic problem is the asymptotic series of eigenfunction $\psi(x)$ for large complex $x$.

Our starting point about the solution of (\ref{schro-eq}) is the Floquet
theory. There are two linearly independent basic solutions to
(\ref{schro-eq}), denoted as $\psi_1(x), \psi_2(x)$. As $\psi_1(x+T)
$ and $\psi_2(x+T)$ also satisfy the equation, therefore they must
be linear combinations of the basic solutions,
\begin{eqnarray}
\left( \begin{matrix}
\psi_1(x+T)  \cr \psi_2(x+T)
\end{matrix}\right)
&=&
M
\left( \begin{matrix}
\psi_1(x)  \cr \psi_2(x)
\end{matrix}\right).\label{periodshift}
\end{eqnarray}
The $2\times2$ nonsingular matrix $M$ does not depend on the base
point $x$, it is called the {\em monodromy matrix}.
The Wronskian of $\psi_1,\psi_2$ are constant, so
we have $\m{det}M=1$. Therefore the two eigenvalues of $M$ can be
written as $e^{\pm i\nu T}$, they are called the Floquet
multipliers. The {\em Floquet exponent} $\nu$ is a function of the
eigenvalue and couplings of the potential, $\nu=\nu(\lambda,g)$. In
quantum physics $\nu$ is called the quasimomentum, and $\lambda$ is
(minus of) the energy, stable solution exists only for real $\nu$.
It is a principle problem to find the dispersion relation
$\lambda(\nu,g)$ which is the spectral solution of (\ref{schro-eq}).
A commonly used method to determine the relation of $\nu$ and
$\lambda$ is Hill's method using the infinite determinant. For most
periodic potentials $u(x)$, when the parameters take generic value it
is impossible to write down an explicit analytical solution.
However, it is possible to obtain asymptotic solutions. If the
leading order term and the small expansion parameter are known, we
can derive the subleading terms from the relation obtained from the
Hill's determinant.

This problem has been a classical topic in differential equation and
quantum theory. However, among the literature we have looked at, it seems that there are some gaps on
this topic. The Floquet theory introduced above can be referred as
{\em classical Floquet theory} as it is a well understood topic for
the case of real singly-periodic potential. If the potential is a periodic
function of more general type, for example an elliptic function with multiple periods, is
there an analogous theory for each period? The elliptic functions are meromorphic function in the complex plane, it is very different from real functions in two periodic dimensions.
The classical Floquet theory is not guaranteed applicable for elliptic potentials.
The precise form of Floquet theory for elliptic potential and its relation to
the spectral solution is still not well understood. Consider the {\em ellipsoidal wave equation} for
example. From the general consideration that, when the kinetic
energy is very large the potential can be treated as small
perturbation, i.e. $\lambda\gg\kappa$ where $\kappa$ is the
characteristic strength of the potential (which means the dominant
one among all $g$ or certain ``average" of all $g$), an asymptotic
spectral solution should exist. Its existence can be inferred also
from the relation of the ellipsoidal wave equation and the Mathieu/Lam\'{e}
equation whose large $\lambda$ spectrum are already
known, see e.g. Refs.~\cite{Wang-Guo, MullerKirstenQUANT, NIST, Langmann2004a, Langmann2004b}. However, it seems such large
energy (weak coupling) asymptotic solution has not been given for the ellipsoidal wave
equation. On the other hand, another asymptotic solution has been
obtained sometime ago which gives the spectrum of
small perturbation at a stationary point $x_*$ of the potential \cite{Muller1966},
i.e. $\lambda=-u(x_*)+\m{perturbation}$, with
$\m{``perturbation''}\ll\kappa$. But for the small energy (strong coupling) asymptotic solution
its connection to the Floquet theory has not been clarified.

In this paper we provide some new results concerning the missing
parts mentioned above. In the Section \ref{largeeigenvalueperturbationgeneral} we show that when the energy is large, the classical Floquet theory is only applicable to the period $2\omega_1$ of the elliptic potential, it gives the corresponding weak coupling dispersion relation. We provide a few
examples, including the ellipsoidal wave equation and the Heun equation in the elliptic form, to demonstrate this. In the Section \ref{doublyperiodicfloquettheory} we study a relation between the small energy spectral solutions and the monodromies of wave function associated to the periods $2\omega_2$ and $2\omega_3$, given in (\ref{exponentmag}) and (\ref{exponentdyon}) respectively. This part involves relations not explained by the classical Floquet theory, reveals that the role of the three periods of elliptic potential are on equal footing but with notable differences. Our main result is that for asymptotic spectral solutions of some elliptic potentials we study in this paper there is a one-to-one correspondence between asymptotic solutions and the monodromy of the wave function along a period. In this paper we demonstrate this fact by their eigenvalues, in the second paper Ref.~\cite{wh1608} we provide further evidences by eigenfunctions of certain typical periodic Schr\"{o}dinger operators.

This paper is motivated by our previous works attempting to examine
in detail a few simple examples of the Gauge/Bethe correspondence,
proposed by Nekrasov and Shatashvili \cite{NS0908}, where the
infrared dynamics of some quantum gauge theories is related to the
spectral problem of stationary Schr\"{o}dinger equation with
periodic potentials. Some results presented in this paper, the relations (\ref{exponentmag}) and (\ref{exponentdyon}), are still
puzzling from the perspective of mathematical theory, albeit they
are supported by solid computation and consistent with results
already known. We hope the results presented here be useful
for further clarification.

\section{Classical Floquet theory and Large Eigenvalue Perturbation}\label{largeeigenvalueperturbationgeneral}

In this section our strategy is to use the {\em classical Floquet theorem} to
compute $\nu(\lambda)$ for large $\lambda$. The eigenvalue relation
$\lambda=\lambda(\nu)$ is the reverse of the Floquet exponent
$\nu(\lambda)$, therefore if we can compute the monodromy of the
wave function along the period then we obtain the eigenvalue
expansion. For elliptic potential it raises the question which period among $2\omega_1, 2\omega_2, 2\omega_3=2\omega_1+2\omega_2$ should be responsible for the large $\lambda$ perturbation?
In this section we show by some examples that large $\lambda$ perturbation is associated the the period $2\omega_1$.

\subsection{Large eigenvalue perturbation}

We use a version of WKB perturbation to perform the computation, the large eigenvalue perturbation has been used in spectral analysis since the work of G. Borg and the work of H. Hochstadt, and others, see Refs.~\cite{Magnus-Winkler} (Chapter II) and ~\cite{Eastham1973} (Chapter 4). It is also known as a standard way to generate the infinite many KdV Hamiltonian densities since the seminal work of Gardner, Greene, Kruskal and Miura, this is explained in Ref.~\cite{MGK}. Write the wave function as $\psi(x)=\exp(\int^xv(y)dy)$,
substitute the wave function into the Schr\"{o}dinger
equation (\ref{schro-eq}), we get the relation
\begin{equation}
v_x+v^2=u+\lambda.\label{muira-trans}
\end{equation}
We use the notation
$u_x=\p_xu, u_{xx}=\p_x^2u,\cdots$, etc. Therefore in order to find
all possible asymptotic spectral expansions $\lambda(\nu)$ we can
first find all possible asymptotic solutions for $v(x)$ from the relation (\ref{muira-trans}) in the parameter space of $\lambda, g$,
and then check if the integration (\ref{nufrommonodromy}) gives an
asymptotic series. When $\lambda$ is large, $\f{1}{\sqrt{\lambda}}$ is a natural expansion parameter,
then we can expand $v(x)$ by \cite{Magnus-Winkler, Eastham1973, MGK, classintegrable}
\begin{equation}
v(x)=\sqrt{\lambda}+\sum_{\ell=1}^{\infty}\f{v_\ell(x)}{(\sqrt{\lambda})^\ell}.\label{nuexpansion}
\end{equation}
Substitute the expansion back to (\ref{muira-trans}), we can solve
$v_\ell(x)$ order by order, they are given by the KdV Hamiltonian densities
\begin{align}
&v_1=\f{1}{2}u,\qquad
v_2=-\f{1}{4}u_x, \qquad v_3=-\f{1}{8}(u^2-u_{xx}),\nonumber\\
&v_4=\f{1}{16}(2u^2-u_{xx})_x,\qquad
v_5=\f{1}{32}(2u^3+u_x^2+(u_{xxx}-6uu_x)_x),\quad\mbox{etc}.
\end{align}
The integrals $H_{\ell}=\int v_{2\ell-1}dx$ are commutative with respect to the Poisson bracket of KdV hierarchy, they are interpreted as the conserved charges of an integrable system of infinite dimension.
The time evolution equations of $u(x,t)$ generated by the conserved charges, $\p_tu=\lbrace H_{\ell}, u\rbrace$, are infinite number of nonlinear partial differential equations that generalize the KdV equation. These connections are the basic facts of relating the spectral data of a linear system and evolution of a nonlinear system, see \cite{classintegrable}.
The relevance of KdV Hamiltonians to the problem of WKB anaylsis was noticed in
e.g. Ref.~\cite{bow1977}, but in their treatment the KdV formalism was not
really further used as the potential they treated is not periodic.

The formal large parameter expansion of $v(x)$ in (\ref{nuexpansion}), and later in (\ref{nuexpansion2}), are valid for any smooth potential,
if the potential is periodic then this procedure gives concrete results of spectral solution for the linear equation (\ref{schro-eq}). Combined with the Floquet property of periodic potentials, some spectral results including but not limited to those already known (as given by Refs.~\cite{Whittaker-Watson, mclachlan, Erdelyi3, Arscott1964, Magnus-Winkler, Eastham1973, Wang-Guo, MullerKirstenQUANT, NIST}) can be easily obtained. The eigenvalue can be derived by the monodromy relation alone, without the eigenfunction at this moment,
\begin{equation}
i\nu T=\int_{x_0}^{x_0+T}v(x)dx.\label{nufrommonodromy}
\end{equation}
In this relation multiplying the period $T$ is required by the classical Floquet theorem.
We have picked the positive sign of $e^{\pm i\nu T}$, to obtain the
result for the other sector we just change the sign of $\nu$.

We emphasize that for general periodic potentials there is no known direct
relation with the KdV theory, and the {\em formal} connection to the KdV
theory is only helpful for computation. However, there are some
potentials $u(x)$ with the special choice of coupling strength which
solve some higher order generalized stationary KdV hierarchy
equations associated to the Hamiltonians given by $\int v_{2\ell-1}dx$. These
special potentials include the Lam\'{e} potential and the
Darboux-Treibich-Verdier potential with triangular number coupling
constants, see e.g. Refs.~\cite{novikov, lax, tv}

When $u(x)$ and its derivatives are periodic, we can abandon all terms
of total derivative in $v_\ell$, and especially the even terms do
not contribute, $\int_{x_0}^{x_0+T}v_{2\ell}dx=0$. Denoting the nonzero integrations by
$\varepsilon_\ell=\f{1}{T}\int_{x_0}^{x_0+T}v_{2\ell-1}dx$,
it depends on the parameters of the potential but not on $x_0$. Then from
(\ref{nufrommonodromy}) and (\ref{nuexpansion}) we have the relation
\begin{equation}
i\nu=\sqrt{\lambda}+\sum_{\ell=1}^{\infty}\f{\varepsilon_\ell}{(\sqrt{\lambda})^{2\ell-1}}.
\end{equation}
For many periodic potentials it is very straightforward to
explicitly compute $\varepsilon_\ell$ because $v_{2\ell-1}$ are
polynomials of $u(x)$ and its derivatives. In this way we obtain the
asymptotic expansion of $\nu(\lambda)$. Reverse the relation, we get
the asymptotic expansion for the eigenvalue,
\begin{equation}
\lambda=-\nu^2+\sum_{l=0}^{\infty}\f{\lambda_l}{\nu^{2l}},\label{lambdanu}
\end{equation}
with $\lambda_0=-2\varepsilon_1,
\lambda_1=\varepsilon_1^2+2\varepsilon_2,
\lambda_2=-2(\varepsilon_1^3+3\varepsilon_1\varepsilon_2+\varepsilon_3),\cdots$.
The large $\lambda$ (therefore large $\nu^2$) expansion
(\ref{lambdanu}) is actually degenerate for $\pm\nu$.

In the following part of this section we apply this method to Schr\"{o}dinger operators with elliptic potentials to show that the periodic $T$ has to be $T=2\omega_1$.

\subsection{Application to elliptic potentials}\label{largeeigenvalueperturbationexamples}

\subsubsection{Hill's equation}

We start with the Hill's equation as an example, the results obtained here is useful for our study of elliptic potentials because it is an important consistence evidence that under certain limit the formulae obtained for elliptic potentials correctly reduce to the formulae of the Hill's equation. The examples in this subsection confirm our conclusion of this section: the relation (\ref{nufrommonodromy}) with $T=2\omega_1$ leads to the spectra for the ellipsoidal wave equation (\ref{ellipsoidalWeiereigenvalue}) and (\ref{ellipsoidalJaceigenvalue}) with correct limit to (\ref{largelambdaWH}), but applying the relation (\ref{nufrommonodromy}) for $T=2\omega_2$ or $T=2\omega_3$ would lead to wrong solutions.

The Hill's equation often refers to equation of the form (\ref{schro-eq}) with
a general single real periodic potential. By the Fourier expansion the
potential can be represented by a trigonometric polynomial,
\begin{equation}
u(x)=\sum_{n=1}^{\infty}2\theta_n\cos2nx,
\end{equation}
the period is $\pi$.
The coupling constants are $\theta_n$, in some cases they may be truncated to a finite subset if the
approximation is valid. The Hill's equation was used in celestial
mechanics to achieve a high-accuracy description of the motion of
moon under the influence of earth and sun. Let us specify to the simple case with $\theta_n=0$ for $n\geqslant3$,
\begin{equation}
u(x)=2\theta_1\cos2x+2\theta_2\cos4x.
\end{equation}
The resulting equation is called the Whittaker-Hill equation. It arises when
we rewrite the 3-dimensional wave equation $\nabla^2W+\widetilde{\chi}^2W=0$ in
the paraboloidal coordinates and apply the separation of variables
method, the wave equation reduces to three identical equations of
Whittaker-Hill type \cite{Arscott1964}.

The integration results
for $\varepsilon_\ell$ are
\begin{equation}
\varepsilon_1=0,\qquad
\varepsilon_2=-\f{1}{4}(\theta_1^2+\theta_2^2),\qquad
\varepsilon_3=\f{1}{8}(2\theta_1^2+8\theta_2^2+3\theta_1^2\theta_2),\quad\mbox{etc},
\end{equation}
and then by (\ref{lambdanu}) we obtain
\begin{align}
\lambda=&-\nu^2-\f{\theta_1^2+\theta_2^2}{2\nu^2}-\f{2\theta_1^2+8\theta_2^2+3\theta_1^2\theta_2}{4\nu^4}\nonumber\\
&-\f{16\theta_1^2+256\theta_2^2+120\theta_1^2\theta_2+5\theta_1^4+40\theta_1^2\theta_2^2+5\theta_2^4}{32\nu^6}+\cdots.\label{largelambdaWH}
\end{align}

\subsubsection{Ellipsoidal wave equation}

Our main concern is the spectral problem with elliptic potentials. The elliptic function we discuss can be represented by the elliptic theta function, Jacobian elliptic functions or Weierstrass elliptic function, see the Appendix \ref{ellipticfunctionconvention} for the convention used in this study.
In the form of Weierstrass elliptic function $\wp(x;2\omega_1,2\omega_2)$, the elliptic potentials have two periods $2\omega_1,2\omega_2$, the periods are independent vectors in the complex plane with the ratio satisfies $\mbox{Im}(\frac{\omega_2}{\omega_1})\ne0$. From $2\omega_1$ and $2\omega_2$ we can make the third period $2\omega_3=2\omega_1+2\omega_2$ which is also needed in the study. Superficially they seem on equal footing regarding the Floquet property.
Nevertheless, it is questionable whether the classical Floquet theory can be directly applied to all
periods in the form (\ref{nufrommonodromy}) or (\ref{mufrommonodromy}). Although there are some discussions directly devoted this problem \cite{Floquet1884, Picard1881, Hermite1885, Arscott-Wright1969, Sleeman-Smith-Wright1984}, a clear connection to the (multiple) asymptotic solutions is absent. As we would show some evidences in the rest of the paper, the classical Floquet theory is still valid for one period $2\omega_1$, but a generalization is needed for other two periods $2\omega_2$ and $2\omega_3$.

If we rewrite the 3-dimensional wave equation $\nabla^2W+\widetilde{\chi}^2W=0$
in the ellipsoidal coordinates, apply the separation of variables
method, then the three identical equations are {\em ellipsoidal wave
equation} \cite{Arscott1964}. Written in the Jacobian form it
is \begin{equation}
\p_z^2\psi(z)-(\Delta k^2\sn^2z+\Omega
k^4\sn^4z)\psi(z)=\Lambda\psi(z),\label{ellipsoidalJacobian}
\end{equation}
where $\Omega\propto\widetilde{\chi}^2$, and $\sn\,z=\sn(z|k^2)$ is the Jacobian
elliptic function with the elliptic modulus $k$, its quarter periods are the
complete elliptic integrals $\mathbf{K}(k^2)$ and $i\mathbf{K}^{\prime}(k^2)=i\mathbf{K}(1-k^2)$.
The periods of $\sn\,z$ are $4\mathbf{K}$ and $2i\mathbf{K}^{\prime}$, the periods of potential are $2\mathbf{K}$ and $2i\mathbf{K}^{\prime}$. In the Weierstrass form it is
\begin{equation}
\p_x^2\psi(x)-(\alpha_1\wp(x)+\alpha_2\wp(x)^2)\psi(x)=\lambda\psi(x),\label{ellipsoidalWeierstrass}
\end{equation}
where $\wp(x)=\wp(x;2\omega_1,2\omega_2)$ is the Weierstrass elliptic
function. The following relations between $x,\wp(x)$ and $z,\sn\,z$ are used,
\begin{equation}
x=\f{z+i\mathbf{K}^{\prime}}{(e_1-e_2)^{1/2}}, \qquad
\wp(x)=e_2+(e_3-e_2)\sn^2z,\label{z-xrelation}
\end{equation}
where
$e_i=\wp(\omega_i)$ and they satisfy $e_1+e_2+e_3=0$. The relation
between half periods is $\mathbf{K}=(e_1-e_2)^{1/2}\omega_1,
i\mathbf{K}^{\prime}=(e_1-e_2)^{1/2}\omega_2$. The nome of the function $\wp(x)$ and
$e_i$ is $q=\exp(2\pi i\f{\omega_2}{\omega_1})=\exp(-2\pi\f{\mathbf{K}^{\prime}}{\mathbf{K}})$,
it is related to the elliptic modulus $k$ by
\begin{equation}
k^2=\f{e_3-e_2}{e_1-e_2}=\f{\vartheta_2^4(q)}{\vartheta_3^4(q)}.\label{kqrelation}
\end{equation}
The
parameters ($\alpha_1,\alpha_2,\lambda$) are related to
($\Delta,\Omega,\Lambda$) by
\begin{equation}
\alpha_1=\Delta-\f{2e_2\Omega}{e_1-e_2},\qquad\alpha_2=\f{\Omega}{e_1-e_2},\qquad
\lambda=(e_1-e_2)\Lambda-e_2\Delta+\f{e_2^2\Omega}{e_1-e_2}. \label{ellipsoidalparamrelat}
\end{equation}

Both Jacobian form and Weierstrass form are useful for our study,
although equivalent but are preferred for computation of different asymptotic solutions.
The Weierstrass form is more suitable for deriving the
large $\lambda$ perturbation given in this section, and the Jacobian
form is more suitable for other two perturbations given in the next
section.

\vspace{3mm}
{\bf The large energy asymptotic solution}

We compute the large energy perturbation with $T=2\omega_1$ in the relation (\ref{nufrommonodromy}). The integrands $v_{2\ell-1}$ contain higher powers of $\wp(x)$ and
$\wp^{'}(x)$, where the prime denotes $\p_x$, they can be simplified
using relations derived from the basic relation
$\wp^{'}(x)^2=4\wp^3(x)-g_2\wp(x)-g_3$. The simplified integrands, after
discarding total derivative terms, take the form
$p_0(g_2,g_3)+p_1(g_2,g_3)\wp(x)$ which is ready for integration, where $p_0, p_1$ are polynomial functions with arguments $g_2, g_3$.
The integration results for $\varepsilon_\ell$ are
\begin{subequations}
\begin{align}
&\varepsilon_1=-\f{1}{2}\alpha_1\zeta_1+\f{1}{24}\alpha_2g_2,\\
&\varepsilon_2=-\f{1}{96}\alpha_1^2g_2+\f{1}{80}\alpha_1\alpha_2(3g_2\zeta_1-2g_3)+\f{1}{2688}\alpha_2^2(48g_3\zeta_1-5g_2^2),\quad \mbox{etc},
\end{align}
\end{subequations}
where $\zeta_1$ is defined by the Weierstrass zeta function $\zeta_1=\f{\zeta(\omega_1)}{\omega_1}$, the modular invariants $g_2, g_3$ are given by $g_2=-4(e_1e_2+e_1e_3+e_2e_3),
g_3=4e_1e_2e_3$. They also can be rewritten in terms of the Eisenstein series $E_2, E_4, E_6$, or in terms of the theta constants $\vartheta_r(q), =1,2,3,4$. We denote the Floquet exponent
of wave function in (\ref{ellipsoidalWeierstrass}) as $\nu$,
i.e. $\psi(x+2\omega_1)=\exp(i2\nu\omega_1)\psi(x)$, then the
asymptotical expansion for $\lambda$ is
\begin{align}
\lambda=&-\nu^2+\f{1}{12}(12\alpha_1\zeta_1-\alpha_2g_2)+\f{1}{5040\nu^2}[105\alpha_1^2(12\zeta_1^2-g_2)
+84\alpha_1\alpha_2(2g_2\zeta_1-3g_3)\nonumber\\
&+10\alpha_2^2(18g_3\zeta_1-g_2^2)]+\mathcal{O}(\f{1}{\nu^4}).\label{ellipsoidalWeiereigenvalue}
\end{align}

The eigenvalue $\Lambda$ for the equation in Jacobian form can be
transformed from $\lambda$. However, the definition for the Floquet
exponent differs \cite{wh1108}. We use $\mu$ to denote the Floquet exponent of
wave function in Jacobian form (\ref{ellipsoidalJacobian}), i.e.
$\psi(z+2\mathbf{K})=\exp(i2\mu\mathbf{K})\psi(z)$. Shifting $x$ by $2\omega_1$ is the
same as shifting $z$ by $2\mathbf{K}$, therefore the phases should be the
same, $\nu\omega_1=\mu\mathbf{K}$, therefore we have
$\nu=(e_1-e_2)^{1/2}\mu$. Taking into account the relation in
(\ref{ellipsoidalparamrelat}),  the relation of $q$ and $k$, we
obtain
\begin{align}
\Lambda=&-\mu^2-[\f{\Delta}{2}k^2+\f{\Delta+6\Omega}{16}k^4+\f{\Delta+2\Omega}{32}k^6+\f{41\Delta+70\Omega}{2048}k^8+\cdots]\nonumber\\
&-\f{1}{\mu^2}(\f{\Delta^2}{32}k^4+\f{\Delta\Omega}{16}k^6-\f{\Delta^2-8\Delta\Omega-136\Omega^2}{4096}k^8-\f{(\Delta-4\Omega)(\Delta+2\Omega)}{4096}k^{10}+\cdots)\nonumber\\
&+\cdots.\label{ellipsoidalJaceigenvalue}
\end{align}
This expression can also be derived by directly applying the large $\Lambda$ perturbation for the equation in the form (\ref{ellipsoidalJacobian}), the relation (\ref{nufrommonodromy}) should be replaced by
\begin{equation}
i\mu T=\int_{z_0}^{z_0+T}v(z)dz.\label{mufrommonodromy}
\end{equation}
with $T=2\mathbf{K}$. The integrand contains terms $\sn^{2m}z, m\in\mathbb{Z}_+$, the integrals are explained in the Appendix \ref{appendixcontourIJ}.

Taking the limit $\Omega\to0$ we recover the results for the
Lam\'{e} equation, already treated in Ref.~\cite{wh1401}, see also Ref.~\cite{wh1108} and references therein. The Lam\'{e} equation comes from
the same procedure of solving the Laplace equation $\nabla^2W=0$ in
the ellipsoidal coordinates. Taking the limit $k^2\to0$ while keeping
$\Delta k^2\to-4\theta_1-16\theta_2,\Omega k^4\to16\theta_2$ we
recover the result for the Whittaker-Hill equation. Taking a further
limit $\theta_2\to 0$ we recover the result for the Mathieu equation.

It is easy to examine that if the period $2\omega_2$ or $2\omega_3$ is wrongly used in large eigenvalue perturbation relation (\ref{nufrommonodromy}), the eigenvalue obtained could not reduce to the eigenvalue  of Whittaker-Hill equation (\ref{largelambdaWH}).

\subsubsection{Heun equation in elliptic form}\label{heunlargeeigen}

A generalization of the Lam\'{e} equation is the Heun equation in the elliptic form.
In the Jacobian form given by G. Darboux \cite{darboux} the equation is
\begin{equation}
\p_z^2\psi(z)-\Bigg(b_0k^2\sn^2z+b_1k^2\f{\cn^2z}{\dn^2z}+b_2\f{1}{\sn^2z}+b_3\f{\dn^2z}{\cn^2z}\Bigg)\psi(z)=\Lambda\psi(z).\label{HeunDarboux}
\end{equation}
In the Weierstrass form it is
\begin{equation}
\p_x^2\psi(x)-\sum_{s=0}^{3}b_s\wp(x+\omega_s)\psi(x)=\lambda\psi(x),\label{HeunWeierstrass}
\end{equation}
where $\omega_0=0$. The multi-component potential in
(\ref{HeunWeierstrass}) is the so-called Treibich-Verdier potential,
known for its role in the theory of ``elliptic soliton" for KdV
hierarchy \cite{tv}. The spectral solution of this potential is related to the effective action of the deformed N=2 supersymmetric QCD model, in the spirit of Gauge/Bethe
correspondence \cite{NS0908}. The large energy perturbation, computed by the method explained in this section, completes the attempts in Ref.~\cite{wh1306}
where the leading order expansion was examined by another method.

As in the previous example, the equation in the form
(\ref{HeunWeierstrass}) is more suitable for the large $\lambda$
expansion. In the process of computing $\varepsilon_\ell$ we need to
simplify the integrands by some more complicated relations of elliptic functions.
We do not give further explicit expression here.
The conclusion is the same as the cases for the Lam\'{e} equation and the ellipsoidal wave equation, that the period $T$ in the relation (\ref{nufrommonodromy}) for large eigenvalue perturbation has to be $2\omega_1$.

\section{On doubly-periodic Floquet theory}\label{doublyperiodicfloquettheory}

\subsection{Spectral problem for elliptic potentials}

We have shown that for elliptic potentials the large $\lambda$
asymptotic solution is always related to the monodromy along the
periodic $2\omega_1$. So what is the role for $2\omega_2$ and
$2\omega_3$? This question is related to the generalized Floquet
theory for elliptic function, the so-called {\em doubly-periodic
Floquet theory}, which only has been occasionally discussed during
the past, e.g. in Refs.~\cite{Floquet1884, Picard1881, Hermite1885,
Arscott-Wright1969, Sleeman-Smith-Wright1984}.
Some new features due to the complex nature of
the elliptic function arise, make the extension nontrivial.
Although it is not a very well understood subject,
in this paper we would use this term for this open problem. Among the limited
results that already exist on this topic, it seems that there is not
an explicit statement about the relation of monodromy along
$2\omega_2$, $2\omega_3$ and the spectrum of the equation. In this
section we give a few examples to show that the monodromy of $v(x)$
along $2\omega_2$ and $2\omega_3$ indeed play a role in the spectral
problem, they are related to two other asymptotic solutions that
differ from (\ref{ellipsoidalJaceigenvalue}) given above.

Therefore the problem we are trying to answer is related to the
complete characterization of all asymptotic spectra for elliptic
potentials. For such a Schr\"{o}dinger operator
the spectral solution $\lambda$ is controlled by the characteristic
coupling strength of potential $\kappa$, or more precisely by the
ratio $\f{\nu}{\kappa}$, often it has no analytical expression. How
does the relation $\lambda(\nu,\kappa)$ vary when we turn the value
of $\f{\nu}{\kappa}$? When $\lambda(\nu,\kappa)$ can be represented
by an asymptotic series? The answer is not obvious. In the
literature it is even not systematically studied how many asymptotic
solutions there are for an elliptic potential.

It is necessary to explain the meaning of ``spectrum" for a complex
potential. The elliptic potentials are meromorphic function defined
on the complex plane, therefore, they are not the most suitable
examples for quantum mechanics. Instead their appearance in quantum
field theory looks more natural, where the complex valued spectrum
of Schr\"{o}dinger operator is explained in a very different way.
Indeed, in the context of Gauge/Bethe correspondence \cite{NS0908}
the spectral solution of elliptic potentials nicely fits into the
theory of 4-dimensional quantum gauge theory. For the Lam\'{e}
potential, due to its connection with a typical Seiberg-Witten gauge
theory model \cite{SW9407, SW9408}, the idea of using elliptic curve
is very helpful for the analysis. Translate the property of the elliptic curve to the property of corresponding elliptic potential,
we are lead to a physical explanation why there is a one-to-one
correspondence between the asymptotic solutions and the monodromy of
wave function along $2\omega_i, i=1,2,3$ \cite{wh1108}. Upon a
careful examination, the complete spectral solutions are precisely
related to nonperturbative and duality properties of the low energy
effective gauge theory. Another related context for the elliptic
potential is the algebraic integrable theory, see e.g. Ref.~
\cite{classintegrable}, albeit neither the questions mentioned above
have been seriously addressed there.

\subsection{Lam\'{e} equation }

The Lam\'{e} potential is the first example that motivates us to
revise the doubly-periodic Floquet theory from a new perspective. It is
$u(x)=\Delta\wp(x)$ in the Weierstrass form, or $u(z)=\Delta
k^2\sn^2z$ in the Jacobian form. The results are already given in Ref.~
\cite{wh1108}, and can be recovered from the case of Ellipsoidal wave equation treated in the next subsection. So here we do not repeat details of the Lam\'{e} potential, only briefly review the result to give a general picture about the (conjectural) Floquet property for elliptic potentials.

The first fact is about the stationary points of the potential.
There are three stationary points for the potential given by the
solutions of $\p_x\wp(x)=0$, they are at
$x_*=\omega_1,\omega_2,\omega_3$ where we have $u(x_*)=e_1\Delta,
e_2\Delta, e_3\Delta$. In the Jacobian form the three stationary
points are given by the solutions of $\p_z\sn^2z=0$, they are at
$z_*=\mathbf{K}+i\mathbf{K}^{\prime},0$ and $\mathbf{K}$, where $u(z_*)=\Delta, 0$ and $\Delta k^2$. The
information about these stationary points does not tell us what are
the possible asymptotic solutions, the following facts entirely come
from computation \cite{wh1108}.

It turns out that each stationary point is associated to an
asymptotic expansion for $\lambda$. The stationary point at
$x_*=\omega_1$ (i.e. at $z_*=\mathbf{K}+i\mathbf{K}^{\prime}$) is associated to the large
eigenvalue (or weak coupling) solution. The equation in the Weierstrass form is better
for computation. The leading order energy comes from the
quasimomentum, $\lambda=-\nu^2+\cdots$, the potential can be treated
as small perturbation, therefore we have
$\nu\sim\sqrt{-\lambda}\gg\kappa\sim\Delta$. The relation
$\nu(\lambda)$ is given by the monodromy along the period
$2\omega_1$ as in the formula (\ref{nufrommonodromy}). This is well
described by the classical Floquet theory, the asymptotic solution
can be treated by the method given in the Section \ref{largeeigenvalueperturbationgeneral}.

The other two stationary points are related to two other asymptotic
solutions, the small eigenvalue (or strong coupling) solutions.
The equation in the Jacobian form is better for computation.
In these cases the quasimomentum is small compared to the scale of potential which means
$\mu\ll\kappa\sim\Delta k^2$.
The solution $\Lambda\sim0+\cdots$ (i.e.
$\lambda\sim-e_2\Delta+\cdots$) is a perturbation at $z_*=0$ (i.e.
at $x_*=\omega_2$), here we have $\Lambda\ll\Delta k^2$. The relation $\mu(\Lambda)$ is given
by the monodromy of wave function along the period $2i\mathbf{K}^{\prime}$ (i.e.
$2\omega_2$). A key point is that the naive definition of the
Floquet exponent $\psi(z+2i\mathbf{K}^{\prime})=\exp(-2\mu\mathbf{K}^{\prime})\psi(z)$ is not
right. If we want to produce the correct asymptotic expansion that
is already derived by other method in Ref.~\cite{Muller1966}, then a
modification is needed, the right relation is
$\psi(z+2i\mathbf{K}^{\prime})=\exp(\mu\pi)\psi(z)$. The solution
$\Lambda\sim-\Delta k^2+\cdots$ (i.e $\lambda\sim-e_3\Delta+\cdots$)
is a perturbation at $z_*=\mathbf{K}$ (i.e. at $x_*=\omega_3$). The subleading
terms are denoted by $\widetilde{\Lambda}$, i.e. $\Lambda=-\Delta
k^2+\widetilde{\Lambda}$ with $\widetilde{\Lambda}\ll\Delta
k^2$. Then the relation $\mu(\widetilde{\Lambda})$ is given by the monodromy along the
period $2(\mathbf{K}+i\mathbf{K}^{\prime})$ (i.e. $2\omega_3$). Again the classical Floquet
theory fails, and the correct definition of Floquet exponent is given
by $\psi(z+2\mathbf{K}+2i\mathbf{K}^{\prime})=\exp(\f{\mu\pi}{ik^{\,\prime}})\psi(z)$, with the
complementary module $k^{\,\prime}$ satisfying the relation $k^{\,\prime\,2}+k^2=1$.

While we do not have a mathematical theory to explain why the
monodromies along three periods are in one-to-one correspond with
three asymptotic solutions, nevertheless a physical explanation can be
given \cite{wh1108}. Viewed from the Gauge/Bethe correspondence
\cite{NS0908}, the spectral problem of the Lam\'{e} operator is
roughly the same problem about the low energy effective theory of N=2$^*$
gauge theory model. The monodromies along different periods are
related by electro-magnetic duality of the effective gauge theory,
in the spirit of Seiberg-Witten theory \cite{SW9407, SW9408}. For
the gauge theory model there is an asymptotic expansion in each
duality frame, hence for the Lam\'{e} operator there is an
asymptotic solution related to the monodromy along each period.

We can use Schr\"{o}dinger equations with more general elliptic potentials to test the relations observed for the Lam\'{e} potential.
In the following we present result for the ellipsoidal wave equation as the main example.

\subsection{Ellipsoidal wave equation}\label{ellipsoidaleigenvalue2e3}

The ellipsoidal wave equation is non-Fushian, nevertheless,
its asymptotic solutions are very similar to the Lam\'{e} potential.
Therefore the strong coupling solutions studied in this subsection provides another evidence for the conjectural Floquet theorem for elliptic potentials.
It is not directly related to gauge theory regarding the Gauge/Bethe correspondence,
but it is a special case of a more general elliptic potential that arises in the study of Gauge/Bethe correspondence \cite{whprep}.

The stationary points of the Lam\'{e} potential are also the
stationary points of the potential $u(z)=\Delta k^2\sn^2z+\Omega
k^4\sn^4z$. The monodromy along $2\mathbf{K}$ (i.e. $2\omega_1$) gives the
large $\Lambda$ asymptotic solution, this is the result given in
(\ref{ellipsoidalJaceigenvalue}). In the following we give the
computation details to demonstrate that the monodromies along
$2i\mathbf{K}^{\prime}$ and $2\mathbf{K}+2i\mathbf{K}^{\prime}$ give other asymptotic eigenvalues, one of
them was already obtained by another method \cite{Muller1966}, another one is a new result.

\vspace{3mm}
{\bf The first small energy asymptotic solution}

The equation in the Jacobian form (\ref{ellipsoidalJacobian}) is
more suitable for this asymptotic solution. We assume $\Delta k^2\sn^2z$ is the dominant term of
the potential, i.e. $\kappa=\Delta k^2$, the other term $\Omega
k^4\sn^4z$ is a small perturbation. At the point $z_*=0$, we have
$u(z_*)=0$, therefore $\Lambda$ is the perturbative energy. The
parameters satisfy $\Omega k^4, \Lambda\ll\Delta k^2$. We shall find
the asymptotic expansion for the integrand $v(z)$ from the relation
$v_z+v^2=u+\Lambda$. Now the expansion parameter should be
$\Delta^{\f{1}{2}}k$, or equivalently $\Delta^{\f{1}{2}}$, with
$v(z)$ expanded as
\begin{equation}
v(z)=\sum_{\ell=-1}^{\infty}\f{v_\ell(z)}{(\sqrt{\Delta})^\ell},\label{nuexpansion2}
\end{equation}
then $v_\ell(z)$ can be recursively solved.
The large coupling expansion of $v(z)$ is another version of WKB expansion, see Refs.~\cite{Arscott1964} (Chapter V) and  ~\cite{ErdelyiAsym} (Chapter IV) for discussions.
In this section we show that for elliptic potentials a clearer understanding of the strong coupling spectrum can be achieved if we (1.) first get the expansion of the form (\ref{nuexpansion2}) {\em at the right critical points of the elliptic potential}, in our example the critical points are $z_*=0$ and $z_*=\mathbf{K}$,  (2.) then {\em choose the correct integral contour} for $v(z)$, which combined with the Floquet theory could lead to solutions consistent with known results. The second point is crucial, the analytical properties of elliptic function needs careful treatment when doing contour integrals, only a particular choice of contour is compatible with the Floquet property, see the Appendix \ref{appendixcontourIJ}.

The even terms $v_{2\ell},\ell=0,1,2,\cdots$ are total derivatives, they do not
contribute in the final periodic integration (\ref{exponentmag}).
The nonzero contributions come from $v_{2\ell-1}$, the first few are
\begin{subequations}
\begin{align}
v_{-1}=&k\,\sn\,z,\\
v_1=&-\f{3}{8k\,\sn^3z}+\f{1+k^2+4\Lambda}{8k\,\sn\,z}+\f{1}{8}k\,\sn\,z+\f{1}{2}\Omega k^3\sn^3z,\\
v_3=&-\f{297}{2^7k^3\sn^7z}+\f{139(1+k^2)+76\Lambda}{2^6k^3\sn^5z}-\f{25(1+k^4)+236k^2+104(1+k^2)\Lambda+16\Lambda^2}{2^7k^3\sn^3z}\nonumber\\
&+\f{7(1+k^2)+28\Lambda+12\Omega k^2}{2^6k\,\sn\,z}-\f{k(1+40\Omega(1+k^2)+32\Lambda\Omega)}{2^7}\sn\,z+\cdots.
\end{align}
\end{subequations}

Then we come to the issue of relating the monodromy of wave function along period $2i\mathbf{K}^{\prime}$
to the Floquet exponent $\mu$. According to the classical Floquet
theory, the relation should be $\int v(z)dz=i\mu\times2i\mathbf{K}^{\prime}=-2\mu\mathbf{K}^{\prime}$. Indeed we can use this relation as the definition of the
Floquet exponent. However, the corresponding asymptotic spectral
solution has already been obtained by a different method in Ref.~
\cite{Muller1966}, the result suggests that the classical Floquet
theory cannot be directly applied to the period $2i\mathbf{K}^{\prime}$.  We find the
correct relation between the period integral and the Floquet
exponent is
\begin{equation}
\mu=\f{1}{\pi}\int_{z_0}^{z_0+2i\mathbf{K}^{\prime}}v(z)dz.\label{exponentmag}
\end{equation}
This relation is the same as that for the Lam\'{e}
equation, it leads to the asymptotic solution given in
(\ref{ellipsoidalJaceigenvalue2}) which is the one obtained in
Ref.~\cite{Muller1966}. This is another example showing how the classical Floquet
theory should be generalized for elliptic potentials.

If we denote $\mathrm{I}_m=\int_{z_0}^{z_0+2i\mathbf{K}^{\prime}}\sn^mzdz$, then we
have $\mathrm{I}_m=0$ for $m=1,3,5,\cdots$, and the remaining
non-vanishing $\mathrm{I}_{-m}$ are
\begin{equation}
\mathrm{I}_{-1}=i\pi,\qquad
\mathrm{I}_{-3}=i\pi\f{1+k^2}{2},\qquad
\mathrm{I}_{-5}=i\pi\f{3+2k^2+3k^4}{8},\quad\mbox{etc}.
\end{equation}
They have been used in the previous related computation for the Lam\'{e}
equation in Ref.~\cite{wh1108}, we give more details in the Appendix \ref{appendixcontourIJ}. Reverse the series $\mu=\mu(\Lambda)$
we reproduce the asymptotic expansion given in Ref.~\cite{Muller1966},
\begin{align}
\Lambda=&-i2\Delta^{\f{1}{2}}k\mu-\f{1}{2^3}(1+k^2)(4\mu^2-1)\nonumber\\
&-\f{i}{2^5\Delta^{\f{1}{2}}k}[(1+k^2)^2(4\mu^3-3\mu)-4k^2(4\mu^3-5\mu)]\nonumber\\
&+\f{1}{2^{10}\Delta k^2}[(1+k^2)(1-k^2)^2(80\mu^4-136\mu^2+9)+384\Omega k^4(4\mu^2-1)]\nonumber\\
&+\f{i}{2^{13}\Delta^{\f{3}{2}}k^3}[(1+k^2)^4(528\mu^5-1640\mu^3+405\mu)-24k^2(1+k^2)^2(112\mu^5-360\mu^3+95\mu)\nonumber\\
&+16k^4(144\mu^5-520\mu^3+173\mu)-512\Omega k^4(1+k^2)(4\mu^3-11\mu)]+\cdots.
\label{ellipsoidalJaceigenvalue2}
\end{align}
In this expression we use notations slightly different from that in Ref.~\cite{Muller1966}, in order to keep consistent with our previous paper Ref.~\cite{wh1108} where the difference is explained.

\vspace{3mm}
{\bf The second small energy asymptotic solution}

The second small energy expansion is a perturbation at $z_*=\mathbf{K}$ where $\sn^2z_*=1$, therefore
$u(z_*)=\Delta k^2+\Omega k^4$. Similar to the treatment in Ref.~\cite{wh1108} we set $\Lambda=-\Delta k^2-\Omega
k^4+\widetilde{\Lambda}$, where $\widetilde{\Lambda}$ is the
perturbative energy around the local minimum of potential. The equation becomes
\begin{equation}
\p_z^2\psi(z)+[\Delta k^2\cn^2z+\Omega
k^4\cn^2z(2-\cn^2z)]\psi(z)=\widetilde{\Lambda}\psi(z).
\end{equation}
The parameters satisfy $\Omega k^4, \widetilde{\Lambda}\ll\Delta k^2$,
therefore similar to the case of the previous solution, we choose
$\Delta^{\f{1}{2}}$ as the expansion parameter and expand the
integrand $v(z)$ as
\begin{equation}
v(z)=i\sum_{\ell=-1}^{\infty}\f{v_\ell(z)}{(\sqrt{\Delta})^\ell}.\label{nuexpansion3}
\end{equation}
From the relation $v_z+v^2=u+\widetilde{\Lambda}$, now with $u(z)=-\Delta k^2\cn^2z-\Omega k^4\cn^2z(2-\cn^2z)$, we obtain the expressions for $v_\ell(z)$. The even terms
$v_{2\ell},\ell=0,1,2,\cdots$ are again total derivatives and do not
contribute to the final integration of (\ref{exponentdyon}). Other
$v_{2\ell-1}$ contribute non-vanishing integrations, the first few are
\begin{subequations}
\begin{align}
v_{-1}=&k\,\cn\,z,\\
v_1=&\f{3k^{\,\prime\,2}}{8k\,\cn^3z}+\f{k^2-k^{\,\prime\,2}-4\widetilde{\Lambda}}{8k\,\cn\,z}+\f{k(1+8\Omega k^2)}{8}\cn\,z-\f{1}{2}\Omega k^3\cn^3z,\\
v_3=&-\f{297k^{\prime\,4}}{2^7k^3\cn^7z}-\f{k^{\,\prime\,2}(139(k^2-k^{\,\prime\,2})-76\widetilde{\Lambda})}{2^6k^3\cn^5z}\nonumber\\
&-\f{25(k^4+k^{\prime\,4})-236k^2k^{\,\prime\,2}-104(k^2-k^{\,\prime\,2})\widetilde{\Lambda}+16\widetilde{\Lambda}^2+48\Omega k^4k^{\prime\,4}}{2^7k^3\cn^3z}\nonumber\\
&+\f{7(k^2-k^{\,\prime\,2})-28\widetilde{\Lambda}-4\Omega k^2(2k^2-5k^{\,\prime\,2})+32\widetilde{\Lambda}\Omega k^2}{2^6k\,\cn\,z}\nonumber\\
&-\f{k(1-8\Omega(3k^2-5k^{\,\prime\,2})+32\widetilde{\Lambda}\Omega+64\Omega^2 k^4)}{2^7}\cn\,z+\cdots.
\end{align}
\end{subequations}

Concerning the issue of relating the monodromy of the wave function along period $2\mathbf{K}+2i\mathbf{K}^{\prime}$ and the Floquet exponent $\mu$,
similar to the case of the first small energy asymptotic
expansion, the classical Floquet theory is invalid. Although the
corresponding asymptotic expansion presented below in
(\ref{ellipsoidalJaceigenvalue3}) has not been given in other
literature, there is the requirement of consistent with other known
results. We find the correct relation is given by
\begin{equation}
\mu=\f{ik^{\,\prime}}{\pi}\int_{z_0}^{z_0+2\mathbf{K}+2i\mathbf{K}^{\prime}}v(z)dz.\label{exponentdyon}
\end{equation}
This relation gives the asymptotic expansion
(\ref{ellipsoidalJaceigenvalue3}) consistent with all known results,
especially in the limits of $\Omega\to0$ (the Lam\'{e} potential), and
in the limit $\Omega\to0, k\to0$, with $\Delta k^2$ fixed (the Mathieu potential).

The integration formulae for $v_{2\ell-1}$ in this case are denoted by
$\mathrm{J}_m=\int_{z_0}^{z_0+2\mathbf{K}+2i\mathbf{K}^{\prime}}\cn^mzdz$, we have
$\mathrm{J}_m=0$ for $m=1,3,5,\cdots$, and
\begin{equation}
\mathrm{J}_{-1}=-i\pi\f{1}{k^{\,\prime}},\qquad
\mathrm{J}_{-3}=-i\pi\f{1-2k^2}{2k^{\,\prime\,3}},\qquad
\mathrm{J}_{-5}=-i\pi\f{3-8k^2+8k^4}{8k^{'5}},\quad\mbox{etc},
\end{equation}
they have been used in Ref.~\cite{wh1108} too. After getting the asymptotic
series $\mu=\mu(\widetilde{\Lambda})$ we reverse it, the
final asymptotic expansion for the spectral relation
$\Lambda=-\Delta k^2-\Omega k^4+\widetilde{\Lambda}(\mu)$ is
\begin{align}
\Lambda=&-\Delta k^2-\Omega k^4+i2\Delta^{\f{1}{2}}k\mu+\f{1}{2^3}(1-2k^2)(\f{4\mu^2}{k^{\,\prime\,2}}+1)\nonumber\\
&+\f{i}{\Delta^{\f{1}{2}}k}\lbrace\f{1}{2^5}[\f{(1-2k^2)^2}{k^{\,\prime}}(\f{4\mu^3}{k^{\,\prime\,3}}+\f{3\mu}{k^{\,\prime}})+4k^2k^{\,\prime}(\f{4\mu^3}{k^{\,\prime\,3}}+\f{5\mu}{k^{\,\prime}})]+2\Omega k^4\mu\rbrace\nonumber\\
&-\f{1}{\Delta k^2}[\f{1-2k^2}{2^{10}k^{\,\prime\,2}}(\f{80\mu^4}{k^{\prime\,4}}+\f{136\mu^2}{k^{\,\prime\,2}}+9)-\f{3}{8}\Omega k^4(4\mu^2+k^{\,\prime\,2})]\nonumber\\
&-\f{i}{\Delta^{\f{3}{2}}k^3}\lbrace  \f{1}{2^{13}}[\f{(1-2k^2)^4}{k^{\,\prime\,3}}(\f{528\mu^5}{k^{'5}}+\f{1640\mu^3}{k^{\,\prime\,3}}+\f{405\mu}{k^{\,\prime}})+\f{24k^2(1-2k^2)^2}{k^{\,\prime}}(\f{112\mu^5}{k^{'5}}+\f{360\mu^3}{k^{\,\prime\,3}}+\f{95\mu}{k^{\,\prime}})\nonumber\\
&+16k^4k^{\,\prime}(\f{144\mu^5}{k^{'5}}+\f{520\mu^3}{k^{\,\prime\,3}}+\f{173\mu}{k^{\,\prime}})]+\f{\Omega k^4}{2^5k^{\,\prime\,4}}[4(4k^4-6k^2+3)\mu^3+k^{\,\prime\,2}(36k^4-58k^2+25)\mu]\nonumber\\
&+\Omega^2k^8\mu\rbrace+\cdots.
\label{ellipsoidalJaceigenvalue3}
\end{align}
We write the expansion in a form easy to see its connection to the
eigenvalue of Lam\'{e} equation, in the limit $\Omega\to0$. The
$\Omega$-independent part in expansions (\ref{ellipsoidalJaceigenvalue2}) and
(\ref{ellipsoidalJaceigenvalue3}) are related by a simple transformation involving $\mu\to\f{i\mu}{k^{\,\prime}}$ and $k\to\f{ik}{k^{\,\prime}}$, the transformation is interpreted as the monopole-dyon duality of N=2$^*$ gauge theory, this is already explained in Ref.~\cite{wh1108}. In fact, the complete expansion of eigenvalues (\ref{ellipsoidalJaceigenvalue2}) and (\ref{ellipsoidalJaceigenvalue3}) can be related by a transformation which is related to the monopole-dyon duality of N=2 $N_f=4$ super QCD model \cite{whprep}.

The potential of ellipsoidal wave equation actually has more
stationary points $z_*$ given by the solutions of $2\Omega k^2\sn^2z_*+\Delta=0$. Therefore it raises the question if they are associated to other still unknown asymptotic spectral solutions? At the moment we do not have a definite answer to this question. Even new asymptotic
solutions exist, we suspect they are unlikely given by the monodromy along a
period, therefore not in the scope of Floquet theory.

\subsection{Darboux-Treibich-Verdier potential}

The Darboux-Treibich-Verdier potential is another generalization of the Lam\'{e} potential. We have verified that the (postulated) relations (\ref{exponentmag}) and (\ref{exponentdyon}) for doubly-periodic Floquet theory are applicable to periods $2i\mathbf{K}^{\prime}$ and $2\mathbf{K}+2i\mathbf{K}^{\prime}$(i.e. $2\omega_2$ and $2\omega_3$), similar to the other elliptic potentials we have studied in this section.
However, the detail is more complicated and will be presented elsewhere \cite{whprep}.
Below we only explain a particular feature of the Darboux-Treibich-Verdier potential that makes the problem more complicated.

From the result of Ref.~\cite{wh1306} there are six stationary points
$z_*$ for the Darboux-Treibich-Verdier potential given by the
solutions of $\p_zu(z)=0$,  each corresponds to an asymptotic
spectral solution. Four of them are related to the large
$\lambda$ asymptotic solution given by the result presented in the subsection
\ref{heunlargeeigen}. The other two stationary points are at
$\Lambda_*=-u(z_*)\sim\pm[(b_0-b_1)(b_2-b_3)]^{1/2}k+\mathcal{O}(k^2)$,
where $\Lambda_*$ is the same order of the geometric average of the
potential terms $(b_0b_1b_2b_3)^{1/4}k$,  they are related to the
remaining two asymptotic solutions. The corresponding Floquet
exponents are given by their relation to the monodromy as in
formulae (\ref{exponentmag}) and (\ref{exponentdyon}). Then we can
rewrite the eigenvalue as $\Lambda=\Lambda_*+\delta$ where $\delta$
is the small perturbation around the local minimum, $\delta\ll
\Lambda_*$. Now the problem is to find a proper expansion for the
integrand $v(z)$ suitable for integration from the relation
\begin{equation}
v_z+v^2=\delta+\Lambda_*+b_0k^2\sn^2z+b_1k^2\f{\cn^2z}{\dn^2z}+b_2\f{1}{\sn^2z}+b_3\f{\dn^2z}{\cn^2z}.
\end{equation}
In this case the potential is not dominated by a single term,
instead every term equally contributes to the potential, moreover,
$\Lambda_*$ is a constant of the same order of the ``averaged"
potential. This feature is different from the Lam\'{e} equation and the ellipsoidal wave equation.
It needs more space to explain the details about the proper expansion for $v(z)$ in this case, the results would be given in \cite{whprep}, along with some other related issues.
There it would be more transparent that the choice of the integral contours in the Appendix \ref{appendixcontourIJ} is unavoidable for elliptic potentials.

\section{Conclusion}

The Floquet theory impose strong constraint on the solution for Schr\"{o}dinger equation with periodic potential. The classical Floquet theory for real singly-period potential is well understood.
But the classical Floquet theory does not provide a complete treatment for potentials of elliptic
function, the precise relation of multiple periods $2\omega_1, 2\omega_2, 2\omega_3$ and
spectral theory is unexplained.

We studied this problem for the Lam\'{e} potential in Ref.~\cite{wh1108}, and related
monodromy along all periods to all possible asymptotic solutions.
In this paper we extended our previous work by studying more general elliptic potentials, the ellipsoidal wave equation is the main example. We notice the large eigenvalue perturbation solution is related to the monodromy along the period $2\omega_1$. For small eigenvalue perturbation solutions,
we propose the relations (\ref{exponentmag}) and (\ref{exponentdyon}) to reproduce an already known solution (\ref{ellipsoidalJaceigenvalue2}) and to obtain a new
solution (\ref{ellipsoidalJaceigenvalue3}) .
The connection to the previous study \cite{Floquet1884, Picard1881,
Hermite1885, Arscott-Wright1969, Sleeman-Smith-Wright1984}, if there is any, remains unclear.

In retrospect, the doubly-periodic Floquet theory is the complete theory, even for a singly-periodic potential it gives a more complete explanation.
For example, the Mathieu equation with potential $u(x)=2\theta_1\cos2x$ has three asymptotic spectral solutions \cite{wh1108}. The large $\lambda$ expansion is explained in the context of singly-periodic Floquet theory as in the Sections \ref{largeeigenvalueperturbationgeneral}, but the other two asymptotic expansions lack such an explanation. Now we know that the other two expansions are special limit of the corresponding asymptotic solutions of the equations with elliptic potential given in the Section \ref{doublyperiodicfloquettheory}. In the limit involving $k\to0$ the quarter period $i\mathbf{K}^{\prime}\to i\infty$, therefore we lose the trace of periods $2i\mathbf{K}^{\prime}$ and $2\mathbf{K}+2i\mathbf{K}^{\prime}$.

\appendix

\section{Some formulae for elliptic functions}\label{ellipticfunctionconvention}

In this Appendix we normalize the convention for some elliptic function formulae useful in this paper and the subsequent paper \cite{wh1608}. The references are \cite{Whittaker-Watson, Erdelyi3, Wang-Guo, NIST, tableofintegrals}.

All the elliptic functions in our study can be expressed in terms of the Jacobian elliptic theta functions. The theta functions are represented by the series
\begin{align}
\vartheta_1(\chi,q)&=-iq^{\f{1}{8}}\sum_{n=-\infty}^{\infty}(-1)^nq^{\f{n(n+1)}{2}}e^{i(2n+1)\chi},\quad \vartheta_2(\chi,q)=q^{\f{1}{8}}\sum_{n=-\infty}^{\infty}q^{\f{n(n+1)}{2}}e^{i(2n+1)\chi},\\
\vartheta_3(\chi,q)&=\sum_{n=-\infty}^{\infty}q^{\f{n^2}{2}}e^{2in\chi},\hspace{3.6cm} \vartheta_4(\chi,q)=\sum_{n=-\infty}^{\infty}(-1)^nq^{\f{n^2}{2}}e^{2in\chi}.
\end{align}
The theta constants are
$\vartheta_r=\vartheta_r(q)=\vartheta_r(0,q)$, and the $\chi$-derivatives of theta functions are
$\vartheta_r^{'}=\vartheta_r^{'}(0)=\p_\chi\vartheta_r(\chi,q)|_{\chi\to0}$ and similar for
higher order derivatives, where $r=1,2,3,4$.

The Weierstrass elliptic function can be expressed by
\begin{equation}
\wp(x;2\omega_1,2\omega_2)=(\f{\pi}{2\omega_1})^2\Big(-\p_\chi^2\ln\vartheta_1(\chi,q)+\f{\vartheta_1^{'''}}{3\vartheta_1^{'}}\Big),
\label{Weierstrass2theta}
\end{equation}
with $\chi=\f{\pi
x}{2\omega_1}$ and $q=\exp2\pi i\tau=\exp2\pi
i\f{\omega_2}{\omega_1}$. The constant term on the right-hand side is $\zeta_1=-\f{\pi^2}{12\omega_1^2}\f{\vartheta_1^{'''}}{\vartheta_1^{'}}$.
From the following relations of theta functions,
\begin{align}
&\vartheta_1(\chi+\f{1}{2}\pi,q)=\vartheta_2(\chi,q),\hspace{2cm}
\vartheta_1(\chi+\f{1}{2}\pi\tau,q)=iq^{-\f{1}{8}}e^{-i\chi}\vartheta_4(\chi,q),\\
&\vartheta_1(\chi+\f{1}{2}\pi+\f{1}{2}\pi\tau,q)=q^{-\f{1}{8}}e^{-i\chi}\vartheta_3(\chi,q),
\label{thetashifts}
\end{align}
the periodicity $\wp(x+2\omega_i;2\omega_1,2\omega_2)=\wp(x;2\omega_1,2\omega_2)$, with $i=1,2,3$,
is referred, and the functions $\wp(x+\omega_i;2\omega_1,2\omega_2)$ can be expressed in the same form as in (\ref{Weierstrass2theta}) with $\vartheta_1(\chi,q)$ in the logarithm substituted by $\vartheta_2(\chi,q),\vartheta_4(\chi,q),\vartheta_3(\chi,q)$, respectively.
There is a set of familiar relations about $e_i(q)$ and $\vartheta_r(q)$,
\begin{equation}
e_1=\f{\pi^2}{12\omega_1^2}(-\vartheta_2^4+2\vartheta_3^4),\quad
e_2=\f{\pi^2}{12\omega_1^2}(-\vartheta_2^4-\vartheta_3^4),\quad
e_3=\f{\pi^2}{12\omega_1^2}(2\vartheta_2^4-\vartheta_3^4).\label{eibytheta1}
\end{equation}
The modular invariants $g_2, g_3$ are expressed by theta constants through $g_2=-4(e_1e_2+e_1e_3+e_2e_3),
g_3=4e_1e_2e_3$, the Eisenstein series are given by $\zeta_1=\f{1}{3}(\f{\pi}{2\omega_1})^2E_2, g_2=\f{4}{3}(\f{\pi}{2\omega_1})^4E_4,g_3=\f{8}{27}(\f{\pi}{2\omega_1})^6E_6$.

In the formula (\ref{Weierstrass2theta}), we have the value $e_i$ on the left-hand side by setting $x=\omega_1,\omega_2,\omega_3$.
On the right-hand side the corresponding expressions can be simplified by the fact  $\vartheta_1^{'}\ne0,
\vartheta_2^{'}=\vartheta_3^{'}=\vartheta_4^{'}=0$. We get another set of
useful relations expressing $e_i$ in terms of theta functions,
\begin{equation}
e_1=(\f{\pi}{2\omega_1})^2(-\f{\vartheta_2^{''}}{\vartheta_2}+\f{\vartheta_1^{'''}}{3\vartheta_1^{'}}),\quad
e_2=(\f{\pi}{2\omega_1})^2(-\f{\vartheta_4^{''}}{\vartheta_4}+\f{\vartheta_1^{'''}}{3\vartheta_1^{'}}),\quad
e_3=(\f{\pi}{2\omega_1})^2(-\f{\vartheta_3^{''}}{\vartheta_3}+\f{\vartheta_1^{'''}}{3\vartheta_1^{'}}).\label{eibytheta2}
\end{equation}
Define the differential operator $\mathcal{D}=q\p_q=\f{1}{2\pi i}\p_\tau$. Using the
heat equation
$i\pi\p_\chi^2\vartheta_r(\chi,q)+4\p_{\tau}\vartheta_r(\chi,q)=0$, we
have $-\f{\vartheta_r^{''}}{\vartheta_r}=2\mathcal{D}\ln\vartheta_r^4$.
Then the relations (\ref{eibytheta2}) is equivalent to the following
relations,
\begin{equation}
\mathcal{D}\ln\vartheta_2^4=\f{2\omega_1^2}{\pi^2}(\zeta_1+e_1),\quad
\mathcal{D}\ln\vartheta_4^4=\f{2\omega_1^2}{\pi^2}(\zeta_1+e_2),\quad
\mathcal{D}\ln\vartheta_3^4=\f{2\omega_1^2}{\pi^2}(\zeta_1+e_3).\label{qderiveoftheta}
\end{equation}
The last set of identities we need are the relations of the complete
elliptic integrals $\mathbf{K}, \mathbf{E}$ and the theta constants,
\begin{equation}
\mathbf{K}=\f{\pi}{2}\vartheta_3^2,\qquad \f{\mathbf{E}}{\mathbf{K}}=k^{\,\prime\,2}(1+\f{d\ln \mathbf{K}}{d\ln k}).\label{EandK}
\end{equation}
Taking into account the relation of $k$ and $q$ given in
(\ref{kqrelation}), we have
\begin{equation}
\f{d\ln \mathbf{K}}{d\ln
k}=\f{\mathcal{D}\ln\vartheta_3^4}{\mathcal{D}\ln\vartheta_2^4-\mathcal{D}\ln\vartheta_3^4}=\f{\zeta_1+e_3}{e_1-e_3}.\label{dlnK2ei}
\end{equation}
This can be used to derive the relation (\ref{zetaandEK}).

The Jacobian elliptic functions can be expressed by the theta functions in a similar way,
\begin{equation}
\dn^2(z|k^2)=\p_z^2\ln\vartheta_4(\f{\pi z}{2\mathbf{K}},q)+\f{\mathbf{E}}{\mathbf{K}},\label{JacobiDN2theta}
\end{equation}
from which the expressions for $\sn\,z, \cn\,z$ can be derived.
To verify the consistency of (\ref{Weierstrass2theta}) and (\ref{JacobiDN2theta}), we should use the relation (\ref{z-xrelation}),(\ref{thetashifts}),(\ref{eibytheta1}) and (\ref{EandK}),(\ref{dlnK2ei}).
The Jacobi zeta function is also useful for future study, it is defined by
\begin{equation}
\mbox{zn}(z|k^2)=\p_z\ln\vartheta_4(\f{\pi z}{2\mathbf{K}},q).
\end{equation}

\section{The contour integrals of $\mathrm{I}_m$ and $\mathrm{J}_m$}\label{appendixcontourIJ}

When we use the Jacobian form of the elliptic potential to compute the monodromies, we need to perform integrals $\mathcal{I}_m=\int\sn^mzdz$ and $\mathcal{J}_m=\int\cn^mzdz$, for integers $m$, along three periodic paths of the function $\sn^2z$. Using the following recursion relations we could reduce the problem to computation of the first few integrals for $m=\pm1,\pm2$. The recursion relations for the indefinite integrals are \cite{tableofintegrals}
\begin{eqnarray}
(m+1)\mathcal{I}_m-(m+2)(1+k^2)\mathcal{I}_{m+2}+(m+3)k^2\mathcal{I}_{m+4}-\sn^{m+1}z\,\cn\,z\,\dn\,z=0,\label{recursionforI}\\
(m+1)k^{\,\prime\,2}\mathcal{J}_m-(m+2)(1-2k^2)\mathcal{J}_{m+2}-(m+3)k^2\mathcal{J}_{m+4}+\cn^{m+1}z\,\sn\,z\,\dn\,z=0.
\label{recursionforJ}
\end{eqnarray}
The definite integrals are performed along trajectories bounded inside the period rectangle $[z_0,z_0+2\mathbf{K}]\times[z_0,z_0+2i\mathbf{K}^{\prime}]$ in the $z$-plane.
The integral trajectories are not closed, the endpoints of the trajectories differ by the three period vectors $2\mathbf{K}, 2i\mathbf{K}^{\prime}, 2\mathbf{K}+2i\mathbf{K}^{\prime}$. For $m=\pm 1$ we use the integral formulae \cite{tableofintegrals}
\begin{align}
&\int\sn\,zdz=\f{1}{2k}\ln\f{\dn\,z-k\,\cn\,z}{\dn\,z+k\,\cn\,z},\hspace{1.45cm} \int\f{dz}{\sn\,z}=\f{1}{2}\ln\f{\dn\,z-\cn\,z}{\dn\,z+\cn\,z},\label{integralsjacobisn}\\
&\int\cn\,zdz=-\f{1}{2ik}\ln\f{\dn\,z-ik\,\sn\,z}{\dn\,z+ik\,\sn\,z},\qquad \int\f{dz}{\cn\,z}=-\f{1}{2k^{\,\prime}}\ln\f{\dn\,z-k^{\,\prime}\sn\,z}{\dn\,z+k^{\,\prime}\sn\,z}.\label{integralsjacobicn}
\end{align}
The logarithm expressions in (\ref{integralsjacobisn}) and (\ref{integralsjacobicn}) have branch cuts, we need to choose the correct integral trajectories in the $z$-plane so that the corresponding paths of $\mbox{cd}z=\cn\,z/\dn\,z$ and $\mbox{sd}z=\sn\,z/\dn\,z$ cross the branch cuts in a correct manner.
For the $2i\mathbf{K}^{\prime}$ periodic integrals (\ref{integralsjacobisn}) a trajectory is chosen to ensure the path of $\mbox{cd}z$ does not cross the branch cut $[-\f{1}{k},+\f{1}{k}]$ but does cross the branch cuts $[-1,+1]$. In this way, we get the correct values $\mathrm{I}_{+1}=0, \mathrm{I}_{-1}=i\pi$ used in the Section \ref{doublyperiodicfloquettheory}.
In a similar way, a trajectory is chosen for the $2\mathbf{K}+2i\mathbf{K}^{\prime}$ periodic integrals (\ref{integralsjacobicn}) to obtain $\mathrm{J}_{+1}=0, \mathrm{J}_{-1}=-i\pi\f{1}{k^{\,\prime}}$.

The contour integrals can be explained in another way. We change the variable by $\sn^2z=\xi$, whose inverse $z=\sn^{-1}\sqrt{\xi}$ is a complex version of the Schwarz-Christoffel mapping. A quarter of period rectangle in the $z$-plane is mapped onto half of the $\xi$-plane. By analytical continuation, the whole period rectangle is mapped twice onto the $\xi$-plane, therefore a periodic trajectory in the $z$-plane is mapped to a closed contour in the $\xi$-plane. We denote the contours in the $\xi$-plane by $\alpha,\beta,\gamma$, respectively.

The contour $\alpha$ is related to the large energy perturbation,
the computation in the Jacobian form is carried as follows. The KdV
Hamiltonian densities $v_{2\ell-1}$ only contain $\sn^mz$ for even
$m\in2\mathbb{Z}_+$. Using the recursion relation of $\mathcal{I}_m$,
we only need to perform the integral
\begin{equation}
\int_{z_0}^{z_0+2\mathbf{K}}\sn^2zdz=\f{1}{2}\oint_\alpha\f{\sqrt{\xi}d\xi}{\sqrt{(1-\xi)(1-k^2\xi)}}=\f{\pi}{2}\,_2F_1(\f{1}{2},\f{3}{2},2;k^2)=\f{2(\mathbf{K}-\mathbf{E})}{k^2},
\end{equation}
where $\mathbf{K}=\mathbf{K}(k^2)$ and $\mathbf{E}=\mathbf{E}(k^2)$ are the complete elliptic integrals
of the first and the second
\begin{wrapfigure}{r}{0.5\textwidth}
\vspace{-20pt}
\begin{center}
\includegraphics[width=5cm]{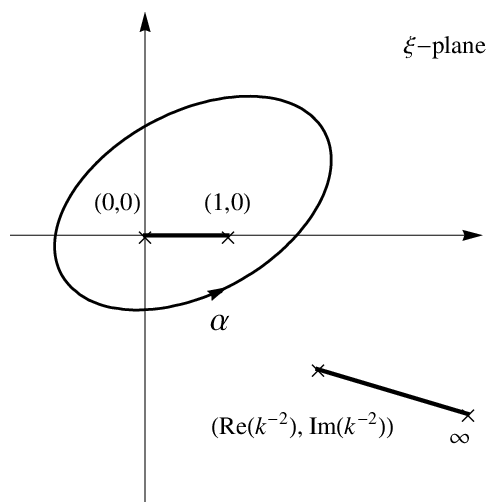}
\end{center}
\vspace{-10pt}
\caption{Integral contour $\alpha$.} \label{alphacontour}
\vspace{-5pt}
\end{wrapfigure}
kind. There are four branch points at $\xi=0,1,\f{1}{k^2}$ and $\infty$,
the branch cuts are between pairs of branch points, as shown in Figure \ref{alphacontour}.
The other necessary integrals are obtained by setting $m=0,2,4,\cdots$ in (\ref{recursionforI}). To compare
with the results in the subsection
\ref{largeeigenvalueperturbationexamples}, we need to use a
relation between elliptic functions,
\begin{equation}
\zeta_1=(e_1-e_2)\f{\mathbf{E}}{\mathbf{K}}-e_1.\label{zetaandEK}
\end{equation}

\begin{figure}[hbt]
\begin{minipage}[t]{0.5\linewidth}
\centering
\includegraphics[width=5cm]{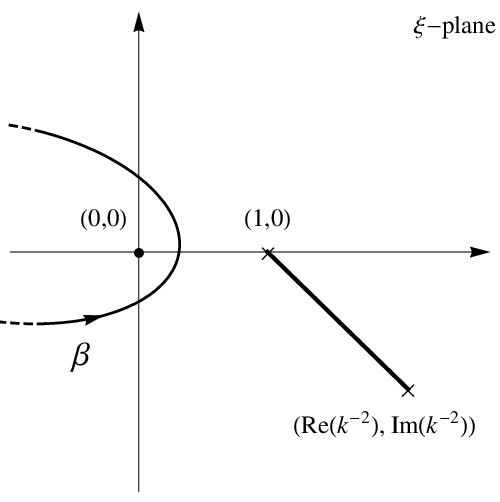}
\caption{Integral contour $\beta$ for $\mathrm{I}_{-1}$.} \label{betacontour}
\end{minipage}%
\begin{minipage}[t]{0.5\linewidth}
\centering
\includegraphics[width=5cm]{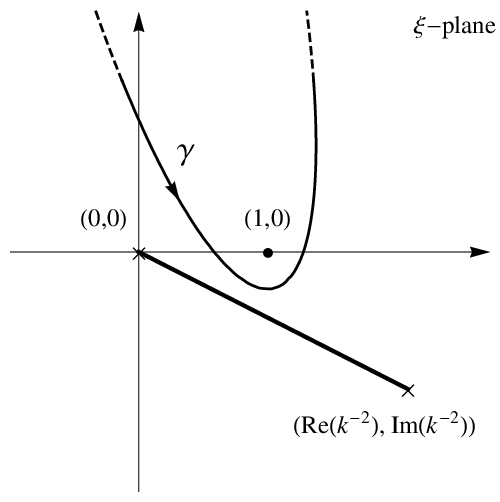}
\caption{Integral contour $\gamma$ for $\mathrm{J}_{-1}$.}
\label{gammacontour}
\end{minipage}
\end{figure}

The contours $\beta$ and $\gamma$ are related to the small energy expansions discussed in the Section \ref{doublyperiodicfloquettheory}, where the definite integrals $\mathrm{I}_m$ and $\mathrm{J}_m$, for odd $m\in2\mathbb{Z}+1$, are used. Using the recursion relations (\ref{recursionforI}), (\ref{recursionforJ}) we only need to perform the integrals of  $\mathrm{I}_{\pm1}$ and $\mathrm{J}_{\pm1}$.
The contours $\beta$ and $\gamma$ are shown in Figure \ref{betacontour} and Figure \ref{gammacontour}, they are chosen to avoid crossing the branch cuts. We draw both contours $\beta$ and $\gamma$ with one side stretched to far away because such periodic trajectories in the $z$-plane typically would pass through the neighbourhood of poles of $\sn\,z$.

The integrals $\mathrm{I}_{\pm1}$ are
\begin{eqnarray}
\mathrm{I}_1=\f{1}{2}\oint_\beta\f{d\xi}{\sqrt{(1-\xi)(1-k^2\xi)}},\qquad \mathrm{I}_{-1}=\f{1}{2}\oint_\beta\f{d\xi}{\xi\sqrt{(1-\xi)(1-k^2\xi)}}.
\end{eqnarray}
There are two branch points at $\xi=1$ and $\xi=\f{1}{k^2}$ for $\mathrm{I}_{\pm1}$, the branch cut is between the branch points. There is a pole at $\xi=0$ for $\mathrm{I}_{-1}$. Then only $\mathrm{I}_{-1}$ receives non-vanishing residue $i\pi$ at $\xi=0$.

The integrals $\mathrm{J}_{\pm1}$ are
\begin{eqnarray}
\mathrm{J}_1=\f{1}{2}\oint_\gamma\f{d\xi}{\sqrt{\xi(1-k^2\xi)}},\qquad \mathrm{J}_{-1}=\f{1}{2}\oint_\gamma\f{d\xi}{(1-\xi)\sqrt{\xi(1-k^2\xi)}}.
\end{eqnarray}
Now there are two branch points at $\xi=0$ and $\xi=\f{1}{k^2}$ for $\mathrm{J}_{\pm1}$, the branch cut is between the branch points. There is a pole at $\xi=1$ for $\mathrm{J}_{-1}$.
Therefore only $\mathrm{J}_{-1}$ receives non-vanishing residue $-i\pi\f{1}{k^{\,\prime}}$ at $\xi=1$.

\section*{Acknowledgments}

I would like to thank  Andrei Mikhailov for reading the paper and
help on improving the presentation. I also thank Chrysostomos
Kalousios for help on computing codes. This work is supported by the
FAPESP No. 2011/21812-8, through IFT-UNESP.

\end{document}